\documentstyle[12pt]{article}
\begin{document}
\begin{titlepage}
\title{On kinematics and dynamics of independent pion emission}
\author{Lajos Di\'osi\\
Institute for Advanced Study,\\ 
Wallotstrasse 19, D-14193 Berlin, Germany\\
and\\
Research Institute for Particle and Nuclear Physics\\ 
H-1525 Budapest 114, POB 49, Hungary\\
{\it E-mail: diosi@rmki.kfki.hu}}
\date{October 4, 1999}
\maketitle
\begin{abstract}
Multiparticle boson states, proposed recently for 'independently' 
emitted pions in heavy ion collisions, are reconsidered in standard second 
quantized formalism and shown to emerge from a simplistic chaotic current 
dynamics. Compact equations relate the density operator, the generating 
functional of multiparticle counts, and the correlator of the external current
to each other. 'Bose-Einstein-condensation' is related to the external pulse. 
A quantum master equation is advocated for future Monte-Carlo simulations.
\end{abstract}
\end{titlepage}
\def\r{\rho}
\def\a{{\hat a}}
\def\k{{\bf k}}\def\kp{{\k^\prime}}
\def\ak{\a_\k}\def\akp{\a_\kp}
\def\Jk{J_\k}\def\Jkp{J_\kp}
\def\f{\varphi}
\def\ket#1{|#1\rangle}
\def\bra#1{\langle#1|}
\def\brak#1#2{\langle#1|#2\rangle}
\def\be{\begin{equation}}
\def\ee{\end{equation}}
\def\bea{\begin{eqnarray}}
\def\eea{\end{eqnarray}}

\section{Introduction}
Multiparticle production in high energy particle collisions is dominated
by classical statistics. Bose-Einstein statistics of pions, nevertheless,
proved to lead to quantum coherence effects which survive in the final 
multiparticle states \cite{GyuKauWil}. The corresponding mechanism can be 
handled within standard quantum statistics. Yet, some recent works 
\cite{Zaj,IMS} continued, mainly for historic reasons as Weiner \cite{Wei} 
points out, to use the "traditional" method of wave function with tedious 
explicit Bose-Einstein-symmetrization. We shall see how standard second 
quantization methods lead to the correct results in a shorter and clearer 
way. I shall discuss the dynamical conditions of Bose-Einsein condensation,
and I outline a master equation suitable for Monte-Carlo simulations. 

For the concentrated study of effects of Bose-statistics, a simple scheme
of 'independent' emission has been proposed 
\cite{IMS,BiaZal.PLB,BiaZal.APS,BiaZal.EJP}. Sec.2 recapitulates the 
features of these multiparticle states, sparing the burden of separate 
'symmetrization'. Sec.3 introduces the generating functional of the 
measured counts \cite{GF} directly in second quantized formalism. 
Introducing chaotic classical currents, advocated e.g. by 
Ref.\cite{GyuKauWil} and used e.g. in \cite{AndPluWei,Ornetal}, I construct 
the simplest quantum dynamics reproducing the corresponding multiparticle 
quantum states (Sec.4). The measured multidetector counts turn out to be 
identical to the corresponding spectral intensities of the effective current 
(Sec.5). I show that the existence of Bose-Einstein condensate imposes
explicit analytic constraints on the intensity and on the spectrum of the 
external effective current (Sec.6). Finally, I generalize the simple 
quantum dynamics and propose a quantum master equation suitable to efficient 
Monte Carlo simulation of the multiparticle density matrix itself (Sec.7).
The Letter concludes with summary (Sec.8).

\section{Independent multiboson states}
When searching for a class of multiparticle density operators $\hat\r$
representing independent bosons, consider first the Gibbs canonical state for
non--interacting bosons at inverse temperature $\beta$. The bosons remain
independent if, formally, we assign different instantaneous temperatures 
$\beta_\k$ to each modes $\k$, i.e. we assume 
$\hat\r\sim\left(-\sum_\k \beta_\k\a_\k^\dagger\a_\k\right)$.
Moreover, the bosons remain independent if we assign different temperatures
to a generic (maybe non--stationary) set of orthogonal modes instead of the
momentum eigenstates. Hence, we arrive at the following class of 'independent
multiboson states' (IMS):
\be
\hat\r=
det(1-e^{-\beta})\exp\left(-\sum_{\k,\kp} \beta_{\kp\k}\akp^\dagger\ak\right)
\ee
where $\beta$ positive matrix. Let us define the 1-particle density matrix 
from the above state:
\be
\r_{\kp\k}=\frac{\bra{\kp}\hat\r\ket{\k}}{\sum_\k\bra{\k}\hat\r\ket{\k}}~.
\ee
Using Eq.~(1), we find the following matrix relation:
\be
e^{-\beta}=\nu\r~~~~(\nu=tre^{-\beta})~.
\ee
The IMS (1) can be rewritten in terms of the 1-particle density matrix $\r$
and the parameter $\nu$ (whose physical interpretation remains a bit 
involved):
\be
\hat\r=
det(1-e^{-\nu\r})\exp\left(\nu\sum_{\k,\kp} 
	\r_{\kp\k}\akp^\dagger\otimes\ak\right)\hat\r_0~.
\ee
This form might give an insight into the kinematics of the particle creation
from the vacuum $\hat\r_0$.

We have to note that the IMS are non-stationary quantum states. Yet, the 
measured quantities $\hat n_\k=\a_\k^\dagger\a_k$ are not sensitive to the 
time evolution of the IMS $\hat\rho$. This will be formulated in the next 
Section.

\section{Generating functional}
We introduce generating functionals for the multiparticle final state 
momentum distributions. A compact heuristic form of definition is the 
following:
\be
G[u]=tr\left(\hat\r\prod_\k(u_\k)^{\hat n_\k}\right)~,
\ee
where $u_\k$ are auxiliary variables and $\hat n_\k=\ak^\dagger\ak$.
If we introduce the diagonal matrix $u$ by 
$u_{\kp\k}=\delta_{\kp\k}u_\k$ then, using the IMS density operator (4),
the generating functional takes the following form:
\be
G[u]=\frac{det(1-\nu\r)}{det(1-\nu u\r)}~.
\ee
The logarithmic generating functional $g=\log G$ can be expressed through
its Taylor--expansion in a transparent way \cite{BiaZal.PLB,BiaZal.EJP}:
\be
g[u]=\sum_{r=1}^\infty \frac{\nu^r}{r}\left(tr(u\r)^r-tr\r^r\right)~.
\ee
For $u_\k\equiv u$, it yields the (logarithmic) multiplicity generating
function
\be
g(u)=\sum_{r=1}^\infty \frac{\nu^r}{r}tr\r^r(u^r-1)
\ee
whose Taylor--coefficients are the combinants (c.f. \cite{GyuKauHeg}).

The derivatives of the generating functionals at $u=0$ yield the
{\it exclusive} distribution/correlation functions. In experiments,
we can easily measure the {\it inclusive} distributions instead, which are
the derivatives at $u=1$ \cite{GF}. To make these derivations 
more convenient, let us substitute $\nu\r$ in the generating functionals
(6-8) by $\nu\rho=\alpha/(1+\alpha)$, where $\alpha$ will be the correlation
matrix of 'currents' of Sec.4:  
\be
G[u]=\frac{1}{det[1-(u-1)\alpha)]}~,
\ee
\be
g[u]=\sum_{r=1}^\infty \frac{\nu^r}{r} tr(u\alpha-\alpha)^r~,
\ee
\be
g(u)=\sum_{r=1}^\infty \frac{\nu^r}{r}tr\r^r(u-1)^r~.
\ee
Comparing these expressions with the Eqs.~(6-8) we see that the inclusive
distributions/correlations will depend on the current correlation matrix
$\alpha$ exactly the same way as the exclusive distributions/correlations
depend on ($\nu$-times) the 1-particle density matrix $\r$.

\section{Dynamics}
Multiparticle production can not be derived from 'first principles'. I can 
certainly not overcome the well-known difficulties. Instead, I present the
simplest quantum dynamics which produces exactly the class (1) of IMS.
I postulate the following effective Hamiltonian:
\be
\hat H=\delta(t)\sum_\k\left(\Jk^\star\ak+\Jk\ak^\dagger\right)~,
\ee
where $\Jk$ denote the Fourier--components of a certain effective external
field $J$ exciting the boson--field. The 'current' $J(x)=J({\bf x})\delta(t)$
is non-zero in the
collision area and we assume that the collision time can be taken infinite 
short. Let us calculate the unitary effect of the Hamiltonian (5) on the 
vacuum:
\be
\ket{\psi_J}=\exp\left(-i\sum_\k(\Jk^\star\ak+\Jk\ak^\dagger)\right)\ket{0}
=\exp\left(-\frac{1}{2}\sum_\k\vert\Jk\vert^2
                    -i\sum_\k\Jk\ak^\dagger\right)\ket{0}~,
\ee
which is otherwise a product coherent state $\prod_\k \otimes\ket{-i\Jk}$.

These final states $\ket{\psi_J}$ are pure states whereas the IMS are mixed 
ones. Obviously, no unitary dynamics can create mixed states from pure ones.
Therefore, I consider unitary dynamics in {\it random} external fields: 
I assume Gaussian distribution for the stochastic fluctuations of the
current $J$. Let the mean values $M[\Jk]$ be always zero. Also we assume that
$M[\Jkp\Jk]\equiv0$, which is equivalent to a random phase for all $\Jk$.
We denote the only non--vanishing correlations by the non-negative
Hermitian matrix $\alpha$:
\be
M[\Jkp\Jk^\star]=\alpha_{\kp\k}~.
\ee
After these preparations, we can define the density operator $\hat\r$ of the
final state as the stochastic mean value of the pure coherent states (13):
\be
\hat\r=M\bigl[\ket{\psi_J}\bra{\psi_j}\bigr]~.
\ee
Substituting the Eq.~(13) and taking the stochastic mean over $J$ of the 
Gaussian correlation (14) we are led directly to
the form (4) of IMS density operators. The 1-particle density matrix $\r$ and
the parameter $\nu$ are related to the correlation matrix $\alpha$ of the
current by easily invertible matrix relations:
\be
\nu\r=\frac{\alpha}{1+\alpha}~,~~~\alpha=\frac{\nu\r}{1-\nu\r}~.
\ee
Measuring the 1-particle density matrix we could, up to the validity of the
model, calculate the structure of the external current. Although such 
measurement is (so far) not completely possible we shall see in Sec.5 that 
the inclusive correlation function gives the modulus of $\alpha$ directly.
It is also seen from Eqs.~(16) that a Gaussian shape, like \cite{BiaZal.EJP}
\be
\r_{\kp,\k}\sim\exp
               \left(-\frac{1}{2\Delta^2}\k_+^2-\frac{1}{2}R^2\k_-^2\right),
~~\k_+=\frac{\k+\kp}{2},~~\k_-=\k-\kp~,
\ee
for the 1-particle density matrix is not compatible with a Gaussian shaped
current correlation matrix $\alpha_{\kp,\k}$ and {\it vice versa}. 

\section{Final state distribution {\it vs.} external current}
The final state distributions in IMS can be directly related to the currents
$J$. The generating functional (5) can conveniently be re-expressed as an
averaged functional over the fluctuating external current $J$:
\be
G[u]=M\left[\exp\left(\sum_\k (u_\k-1)\vert\Jk\vert^2\right)\right]~,
\ee
which is of course equivalent to Eqs. (6) or (9).

The above equation has numerous useful consequences. The multiplicity 
distribution can be written in this form:
\be
p_r=M
\Bigl[\left(\sum\vert J\vert^2\right)^r\exp\left(-\sum\vert J\vert^2\right)
\Bigr]~,
\ee
while the factorial moments take the same form but without the exponential
factor $exp(-\sum\vert J\vert^2)$, i.e.:
\be
F_r=M
\Bigl[\left(\sum\vert J\vert^2\right)^r\Bigr]~.
\ee
This phenomenon also characterizes the differences between the expressions of 
the exclusive and the inclusive distribution functions, respectively:
\bea
f(1,2,\dots,r)=M\Bigl[\vert J_1\vert^2,\vert J_2\vert^2,\dots,\vert J_r\vert^2 
\cases{\exp(-\sum\vert J\vert^2)\Bigr]~~~&(exclusive)\cr
                                \Bigr]   &(inclusive)
}
\eea
as well as of the correlation functions. In particular, the inclusive 
correlation functions take the following form:
\be
C(1,2,\dots,r)=
M\left[\vert J_1\vert^2 \vert J_2\vert^2 \dots \vert J_r\vert^2\right]_c~,
\ee
the inclusive ones would contain the ominous exponential factor, too.
The notation $M[\dots]_c$ means that in the 'expectation value' only the
'connected grafs' are to be taken into the account. In case of Eq.~(22)
it yields $(r-1)!$ 'cycles', i.e. the 'cycle'
$\alpha_{12}\alpha_{23}\dots\alpha_{r1}$ and its variants for permutations
of $2,\dots,r$ \cite{BiaZal.PLB}.

One can easily summarize the main result of this Section as follows. The
counts $n_\k$, measured simultaneously in a collision event, are 
{\it statistically identical} to the corresponding spectral intensities 
$\vert\Jk\vert^2$. Like their distributions, their corresponding moments are 
identical as well:
\be
\langle n_1 n_2 \dots n_r\rangle =
M\left[\vert J_1\vert^2 \vert J_2\vert^2 \dots \vert J_r\vert^2\right].
\ee

\section{Bose-Einstein-condensation}
The IMS class of density operators (1) has a particular asymptotics. The 
'inverse temperature' matrix $\beta$ must be positive. If it were degenerate 
the state (1) would not exist at all. A degenerate $\beta$ can formally be 
interpreted as if the mode of zero eigenvalue became infinite hot. This mode 
is, in fact, becoming more and more populated but the infinite population is
unattainable. Nonetheless, an IMS with almost degenerate $\beta$ would really
be a Bose-Einstein condensate since this only requires a big finite number of 
bosons in a single quantum state. Speculations that the point of degeneracy, 
i.e. the point $\nu=1/\Vert\r\Vert$, is the point of condensation (like in 
thermal Bose--systems) can not be verified for the IMS.

Let us first recapitulate the kinematics of an IMS condensate. The condensate
mode does not interfere with the other modes so we can discuss it separately.
We assume that our IMS is dominated by the condensate mode. The 1-particle
density matrix has the form $\r_{\kp\k}=\f_\kp\f_\k^\dagger$ where $\f_\k$
is the condensate mode's wave function. If we introduce the condensate
absorption operator $\a_c=\sum_\k\f_\k\a_\k$ then, using Eqs.(1-4), the
condensate IMS can be written as a thermal equilibrium state at temperature
$T=-1/\log\nu$:
\be
\hat\r_c=(1-e^{-1/T})\exp\left(-\frac{\a_c^\dagger\a_c}{T}\right)~.
\ee
This state assumes a Hamiltonian $\a_c^\dagger\a_c$ which is not the real 
case, the condensate is not even stationary in general. Yet, the form (24) is
completely proper to calculate characteristics of the state by a thermal 
analogy. For instance, Eq.~(5) yields directly the generating functional in 
the form:
\be
G[u]=\frac{1-e^{-1/T}}{1-\exp(-1/T)\sum_\k u_\k\vert\f_\k\vert^2}~,
\ee
with the canonical thermal multiplicity distribution 
\be
p_n=(1-e^{-1/T})e^{-n/T}
\ee
of mean multiplicity
\be
\langle n\rangle = \frac{1}{e^{n/T}-1}~.
\ee 
Let us observe that approaching the 'condensation point' corresponds to
$T\rightarrow\infty$ and the population of the 'Bose-condensate' increases
to the infinity while it is remaining thermally distributed all the time.

Now I turn to the dynamic conditions for the fluctuating external current
$J$. In the special case of the condensate IMS, the second relation in
Eq.~(16) becomes simply 
$\alpha_{\kp\k}=\langle n\rangle\r_{\kp\k}=\langle n\rangle\f_\kp\f_\k^\star$. 
Recall the definition (14) of $\alpha$ as the current's correlation matrix, 
which yields the following relation: 
\be
M[\Jkp\Jk^\star]=\langle n\rangle \f_\kp\f_\k^\star~.
\ee
Regarding that $M[\Jkp\Jk]$ should vanish by assumption (Sec.4), the Gaussian 
fluctations satisfying the above relation must take the form  
\be
\Jk=z\sqrt{\langle n\rangle}\f_k
\ee
for all $\k$, where $z$ is a random complex number of the standard 
Gauss--distribution $(1/\pi)\exp(-\vert z\vert^2)d^2z$. Taking the
stochastic mean of the modulus square of both sides we obtain:
\be
\vert\Jk\vert^2=\langle n\rangle \vert\f_\k\vert^2~,
\ee
which also leads to
\be
\langle n\rangle=\sum_\k \vert\Jk\vert^2~.
\ee
The Eqs.~(29-31) show the simple way how the pulse of the effective current
$J$ determines the condensate wave function and the mean population. Actually, 
the mean multiplicity is identical to the overall intensity of the current 
pulse (31). The pulse's normalized spectral density is equal to the modulus 
square of the condensate wave function (30). The Eq.~(29) seems, however, to 
be very restrictive since it imposes the same random phase and weight 
simultaneously for all current amplitudes $J_\k$. 

\section{Outlook: master equations}
I outline a possible generalization of the simple dynamics proposed in
Sec.4. Let us replace the Hamiltonian (12) by 
\be
\hat H(t)=g(t)\sum_\k\left(\Jk^\star(t)\ak+\Jk(t)\ak^\dagger\right)~,
\ee%32
where $g(t)$ is a normalized function of characteristic width $\Delta t$, 
controlling the intensity of particle creation. 
The time-dependent currents $J_\k(t)$ are random functions of zero mean;
let their correlator be of non-stationary white-noise type:
\be%33
g(t)M[\Jkp(t^\prime)\Jk^\star(t)]=\delta(t^\prime-t)\alpha_{\kp\k}~.
\ee
In the limit $\Delta t\rightarrow0$, the random dynamics represented by 
Eqs.~(32,33) reduces to the simplistic dynamics (12,14) of Sect.~4.   
The Hamiltonian (32) with the white-noise currents (33) yield the
following master equation  \cite{GorKosSud} for the noise-averaged 
density operator (in interaction picture):
\be
\frac{d\hat\r}{dt}=g
\sum_{\k,\kp}\alpha_{\kp\k}
\left(\akp^\dagger\hat\r\ak-\frac{1}{2}\{\ak\akp^\dagger,\hat\r\}\right)~.
\ee
One solves the master equation with vacuum initial state. 
If $1/\Delta t$ is much greater than the typical pion energy
then, via the relationships (16), the final state will tend to the IMS (1).
For larger $\Delta t$ the simple relations (16) do not hold. Though analytic 
calculations are still possible one can turn to very powerful Monte-Carlo 
methods \cite{Dio88}
developped for Markovian master equations. These MC algorithms will 
yield the density operators of the multiparticle final states without 
"struggling through" \cite{Zaj93} the usual Wigner-function formalism.

\section{Summary}
The aim of the Letter was partly pedagogical. To avoid  
Bose--symmetrization 'by hand', I used standard quantum mechanical 
considerations to construct and to analyze the 'independent multiboson 
states'. I showed how these states emerge from a simplistic version of
chaotic current models and I derived the relationship between the IMS states 
and the correlator of the currents. I briefly recapitulated the generating 
functional representation of multiparticle counts. Beyond methodological 
matters, I found that the 'Bose-Einstein condensate' would be thermally 
populated and the condensation point corresponds to the infinite hot state. 
I restricted my analysis for the IMS of Sec.1 and for the simplest 
chaotic current mechanisms with trivial time-dependences. The simple choice
allows for transparent relationships between the current and the final 
multiparticle state. (Other works, like e.g. Ref.\cite{Ornetal}, incorporate 
a more realistic space-time evolution of the multiparticle source.)
There is, nonetheless, a particular advantage of any underlying dynamics 
whether realistic or not. Usually it allows economic simulation methods for 
the physical quantities of interest. To this end, I proposed a quantum master 
equation known to be suitable to efficient Monte-Carlo simulations.

\section*{Acknowledgement}
I thank Tam\'as Cs\"org\H o and S\'andor Hegyi for useful discussions.

\end{document}